\documentclass[conf]{new-aiaa}

\usepackage{graphicx}
\usepackage[version=4]{mhchem}
\usepackage{siunitx}
\usepackage{longtable,tabularx}
\usepackage{bm}
\usepackage{bbm}
\usepackage{accents}
\usepackage{stackengine}
\usepackage{subcaption}

\newcommand{\R}{\mathbb{R}}

\newcommand{\E}{\operatorname{\mathbb{E}}}
\newcommand{\sN}{\mathsf{N}}

\usepackage{amsthm}
\newtheoremstyle{italicblock}
  {12pt}
  {12pt}
  {\itshape}
  {}
  {\bfseries}
  {}
  {0.5em}
  {}

\theoremstyle{italicblock}

\theoremstyle{italicblock}

\theoremstyle{italicblock}
\newtheorem{theorem}{Theorem}

\theoremstyle{italicblock}

\title{Multi-Spacecraft Predictive Sensor Tasking for Cislunar Space Situational Awareness}

\author{Kento Tomita\footnote{PhD Candidate, Daniel Guggenheim School of Aerospace Engineering, Space Systems Design Lab/Space Systems Optimization Group, 620 Chery St. NW, Atlanta, GA, AIAA Student Member.}, Yuri Shimane\footnote{PhD Candidate, Daniel Guggenheim School of Aerospace Engineering, Space Systems Design Lab/Space Systems Optimization Group, 620 Chery St. NW, Atlanta, GA, AIAA Student Member.}, and Koki Ho\footnote{Associate Professor, Daniel Guggenheim School of Aerospace Engineering, Space Systems Design Lab/Space Systems Optimization Group, 620 Chery St. NW, Atlanta, GA, AIAA Senior Member.}}
\affil{Georgia Institute of Technology, Atlanta, GA, 30332}

\begin{document}

\maketitle

\begin{abstract}
    This paper delves into the predictive sensor tasking algorithm for the multi-observer, multi-target sensor setting, leveraging the Extended Information Filter (EIF). Conventional predictive formulations suffer from the curse of dimensionality due to the dependence of the performance metric on the target-observer assignment history. This paper exploits the EIF's additive structure of measurement information to break the dependence and devises an efficient linear integer programming formulation. We further investigate the resulting formulation to study how the cislunar dynamics expands and shrinks the measurement information, and discuss when the information gain is maximized in relation to the observation space and the uncertainty deformation caused by the dynamics. We numerically demonstrate that the predictive sensor tasking algorithm outperforms the myopic algorithm in two different metrics, depending on the formulation.
\end{abstract}

\section{Introduction}
The increasing complexity and dynamism of space activities necessitate advanced Space Domain Awareness (SDA) capabilities, especially within the translunar and cislunar regions. These regions experience stable observation demands, such as the monitoring of existing assets like the Gateway from the Artemis program\cite{smith2020artemis}, LunaNet service provider satellites\cite{israel2020lunanet}, CLPS satellites in low-lunar orbit \cite{chavers2019nasa}, and several other lunar missions, including those from non-US entities like CNSA’s Queqiao relay satellite at Earth-Moon L2 \cite{gao2019optimization}. Concurrently, there are unpredictable detection and tracking demands, mainly arising from translunar transfers and missions with configurations like JAXA’s SELENE mission\cite{kato2008japanese}, that deploy additional segments upon reaching their destinations. Thus, cislunar SDA systems must be adept at addressing these dynamic requirements.

In the rapidly evolving landscape of the cislunar space, understanding how the Cislunar SDA system can adaptively respond to unforeseen demands and which architectural frameworks for Cislunar SDA demonstrate resilience against unpredictable tracking requirements is paramount. It is in this context that there is an evident need for a predictive sensor tasking algorithm. Such an algorithm should not only be swift and optimal but should also be sufficiently flexible to accommodate shifts in SDA architecture. With these attributes, it would facilitate comprehensive trade studies of the cislunar SDA architectural landscape.

Historically, sensor tasking algorithms for SDA have attracted significant scholarly attention. Initial efforts focused on myopic algorithms \cite{kalandros2005multisensor, kangsheng2006sensor, erwin2010dynamic, williams2013coupling, adurthi2015conjugate} optimized on metrics related to covariance \cite{kalandros2005multisensor} or information-theoretic quantities such as Fisher information gain\cite{kangsheng2006sensor, erwin2010dynamic, williams2013coupling} and mutual information \cite{adurthi2015conjugate}. Notably, the myopic greedy approach has shown consistent performance, sometimes even outperforming reinforcement learning (RL) based strategies \cite{little2020space}. However, the domain of predictive sensor tasking, oriented towards optimizing target-observer assignments over finite time horizons, has faced challenges tied to the curse of dimensionality, as it's influenced by prior target-observer assignments \cite{miller2007new}. As a response, recent approaches have incorporated RL \cite{linares2016dynamic, sunberg2015information, siew2022optimal}, heuristic techniques \cite{frueh2018heuristic, gehly2018sensor, cai2019multisensor, adurthi2020mutual}, and Monte Carlo tree searches \cite{fedeler2022sensor, fedeler2022tasking}. Gualdoni and DeMars' innovative method \cite{gualdoni2020impartial} leveraged the extended information filter (EIF) to create a framework for observation event selection by projecting the observed information at the common evaluation time. This research expands on this methodology.

This paper proposes an efficient sensor tasking algorithm without employing the heuristic algorithms while achieving the near-optimal policy via extended information filter (EIF). The additive structure of EIF's measurement information is harnessed to disconnect the performance metric from prior target-observer assignments, allowing us to devise an efficient linear integer programming formulation. Our investigation delves deep into this formulation, examining the influences of cislunar dynamics on measurement information, identifying conditions under which information gain is optimized, and exploring the implications of dynamics-induced uncertainty. We numerically demonstrate that the predictive sensor tasking algorithm outperforms the myopic algorithm in two different metrics, depending on the formulation.

\section{Background}

\subsection{Filtering Algorithms}

In this subsection we review the extended Kalman filter (EKF) and its information form, also known as the extended information filter (EIF)~\cite{maybeck1982stochastic, schutz2004statistical}. The EKF propagates the covariance matrix whereas the EIF propagates the inverse of the covariance matrix, which is also called the information matrix. Both algorithms approximate the nonlinear state dynamics and measurement model around a nominal state trajectory as a linear system. Although EKF and EIF are well-known algorithms, they are briefly explianed here because they form the basis of our study.
First we review the linearlized state dynamics and measurement model, followed by the prediction step and the update step of the EKF and the EIF.

\subsubsection{System Linearlization}
Let $\bm{x}(t)$ be the state vector and $\bm{\beta}(t)$ be the process noise assumed to be zero-mean Gaussian with covariance matrix $\tilde{Q}(t)$. The nonlinear continuous state dynamics is given by
\begin{equation}
    \begin{aligned}
    \bm{dx}(t) &= f(\bm{x}(t))dt + \bm{d\beta}(t)\\
    \E\left[\bm{\beta}(t)\right] &= \bm{0}, \quad \E\left[\left(\bm{\beta}(t)-\bm{\beta}(t')\right)\left(\bm{\beta}(t)-\bm{\beta}(t')\right)^T\right] = \int_{t'}^{t}\tilde{Q}(\tau)d\tau.
\end{aligned}
\end{equation}

Given a nominal trajectory $\bar{\bm{x}}(t)$, we can linearize the state dynamics around the nominal trajectory. Let $\bm{\xi}(t)=\bm{x}(t)-\bar{\bm{x}}(t)$ be the state error vector. Then, the linearized state error dynamics is given by
\begin{equation}
    \begin{aligned}
    \label{eq:linear-dynamics}
    \bm{\xi}(t_{k}) &= \Phi(t_{k}, t_{k-1})\bm{\xi}(t_{k-1}) + \bm{w}(t_{k-1})\\
    \Phi(t_{k}, t_{k-1}) &= \int_{t_{k-1}}^{t_{k}}Df(\bar{\bm{x}}(\tau))d\tau, \quad
    \Phi(t_{k-1}, t_{k-1}) = I
\end{aligned}
\end{equation}

where 
\begin{equation}
    \E\left[\bm{w}(t_{k-1})\right] = \bm{0}, \quad \E\left[\bm{w}(t_{k-1})\bm{w}(t_{k-1})^T\right] = Q.
\end{equation}

Similarly, we can linearize the measurement model around the nominal trajectory. Let $\bm{y}(t)$ be the measurement vector and $\bm{v}(t)$ be the measurement noise assumed to be zero-mean Gaussian with covariance matrix $R$. Suppose the nonlinear measurement model is given by $\bm{y}(t) = h(\bm{x}(t))+\bar{\bm{v}}(t)$, and the deviation from the nominal measurement is denoted by $\bm{z}(t)=\bm{y}(t) - \bar{\bm{y}}(t)=\bm{y}(t) - h(\bar{\bm{x}}(t))$. Then, the linearized measurement error model is given by
\begin{equation}
    \begin{aligned}
    \label{eq:linear-measurement}
    \bm{z}(t_{k}) &= H(t_k)\bm{\xi}(t_k) + \bm{v}(t_k)\\
    H(t_k) &= Dh(\bar{\bm{x}}(t_k)).
\end{aligned}
\end{equation}
where
\begin{equation}
    \E\left[\bm{v}(t_k)\right] = \bm{0}, \quad \E\left[\bm{v}(t_k)\bm{v}(t_{k})^T\right] = R.
\end{equation}

\subsubsection{Prediction Step}
The prediction step computes the prior distribution of state error before observation, $\bar{\bm{\xi}}(t_{k})$, given the previous posterior distribution by the observation, $\bm{\xi}(t_{k-1})$. The propagation is based on the linear dynamics of Eq. \eqref{eq:linear-dynamics}. Since the state error is assumed to be zero-mean Gaussian and independent of the process noise, the prior distribution at $t_{k}$ is also zero-mean Gaussian with a covariance matrix, denoted by $\bar{\bm{P}}(t_{k})$. The prior covariance matrix $\bar{\bm{P}}(t_{k})$ and its information form $\bar{\bm{\Lambda}}(t_{k})=\bar{\bm{P}}(t_{k})^{-1}$ are computed as the sum of the linear map of $\bm{\xi}(t_{k-1})\sim\sN(\bm{0}, \bm{P}(t_{k-1}))$ and the process noise $\bm{v}(t_{k-1})\sim\sN(\bm{0}, \bm{Q})$ as follows~\cite{rencher2008linear}:
\begin{align}
    \label{eq:ekf-prediction}
    \bar{\bm{P}}(t_{k}) &= \Phi(t_{k}, t_{k-1})\bm{P}(t_{k-1})\Phi(t_{k}, t_{k-1})^T + \bm{Q}\\
    \bar{\bm{\Lambda}}(t_{k}) & = \left(\Phi(t_{k}, t_{k-1})\bm{\Lambda}^{-1}(t_{k-1})\Phi(t_{k}, t_{k-1})^T + \bm{Q}\right)^{-1}.
\end{align}
Another form of the propagation equation for the information matrix is given with the Woodbury identity:
\begin{equation}
    \label{eq:woodbury}
 (A+UCV)^{-1} = A^{-1} - A^{-1}U(C^{-1}+VA^{-1}U)^{-1}VA^{-1}.
\end{equation}
 Let $M=\Phi(t_{k-1}, t_{k})^T\bm{\Lambda}(t_{k-1})\Phi(t_{k-1}, t_{k})$ and substituting $A=M^{-1}$, $C=Q$, and $U=V=I$ to the identity of Eq. \eqref{eq:woodbury}, we obtain
\begin{equation}
    \bar{\bm{\Lambda}}(t_{k}) = M - M\left(Q^{-1}+M\right)^{-1}M.
\end{equation}
The propagation of the information form is computiationally less efficient than that of the covariance matrix because of the inverse operations. However, without the process noise, we obtain the following simpler linear expressions. Note that the backward state transition matrix is used for the information form.
\begin{align}
    \label{eq:eif-prediction-noiseless}
    \bar{\bm{P}}(t_{k}) &= \Phi(t_{k}, t_{k-1})\bm{P}(t_{k-1})\Phi(t_{k}, t_{k-1})^T\\
    \bar{\bm{\Lambda}}(t_{k}) & = \Phi(t_{k-1}, t_{{k}})^T\bm{\Lambda}(t_{k-1})\Phi(t_{k-1}, t_{k})
\end{align}

\subsubsection{Update Step}
The update step computes the posterior distribution of state error after observation, $\bm{\xi}(t_{k})$, given the prior distribution $\bar{\bm{\xi}}(t_{k})$. The update is made to minimize the variance of the conditional distribution of $\bm{\xi}(t_{k})$ given the measurement $\bm{z}(t_{k})$ and the prior distribution $\bar{\bm{\xi}}(t_{k})$. The update in the covariance form is well-known as the Kalman filter update:
\begin{equation}
\begin{split}
    \label{eq:kalman-gain}
    \bm{P}(t_{k}) &= \left(I-\bm{K}(t_{k})H(t_{k})\right)\bar{\bm{P}}(t_{k})\\
    \bm{K}(t_{k}) &= \bar{\bm{P}}(t_{k})H(t_{k})^T\left(H(t_{k})\bar{\bm{P}}(t_{k})H(t_{k})^T + \bm{R}\right)^{-1}
\end{split}
\end{equation}
where $\bm{K}(t_{k})$ is the Kalman gain matrix. 
We can derive the update in the information form by applying the Woodbury identity of Eq. \eqref{eq:woodbury} to the inverse of the covariance matrix in Eq. \eqref{eq:kalman-gain}. Substituting $A^{-1}=\bar{\bm{P}}(t_{k})=\bar{\Lambda}^{-1}(t_k)$, $C^{-1}=R$, $V=U^T=H(t_{k})$ into Eq. \eqref{eq:woodbury}, we obtain
\begin{equation}
    \label{eq:eif-update}
    \bm{\Lambda}(t_{k}) = \bar{\bm{\Lambda}}(t_{k}) + H(t_{k})^T\bm{R}^{-1}H(t_{k}).
\end{equation}
Note that the information form allows the simple additive operation for synthesizing the propagated prior information matrix $\bar{\bm{\Lambda}}(t_{k})$ and the measurement information $H(t_{k})^T\bm{R}^{-1}H(t_{k})$.

\subsubsection{Information Matrix Propagation in EIF}
When we assume no process noise in the EIF, the combined expression of the prediction step and the update step becomes relatively simple. The single step uncertainty propagation of the information form is given by
\begin{equation}
    \label{eq:eif-single-propagation}
    \bm{\Lambda}(t_{k}) = \Phi(t_{k-1}, t_{k})^T\bm{\Lambda}(t_{k-1})\Phi(t_{k-1}, t_{k}) + \Phi(t_{k-1}, t_{k})^TH(t_{k})^T\bm{R}^{-1}H(t_{k})\Phi(t_{k-1}, t_{k})^T.
\end{equation}
We can easily obtain the multi-step propagation by recursively applying Eq. \eqref{eq:eif-single-propagation}. 
\begin{equation}
\begin{split}
    \label{eq:eif-multi-propagation}
    \bm{\Lambda}(t_{L}) &= \Phi(t_{0}, t_{L})^T\bm{\Lambda}(t_{0})\Phi(t_{0}, t_{L}) + \sum_{k=0}^{L-1}\Phi(t_{k}, t_{L})^T H(t_{k})^T\bm{R}^{-1}H(t_{k})\Phi(t_{k}, t_{L})\\
    &\triangleq \mathcal{I}_0(t_{L}) + \sum_{k=0}^{L-1}\mathcal{I}(t_{L}, t_{k}).
\end{split}
\end{equation}
Here $\mathcal{I}_0(t_{L})$ is the initial information matrix propagated to $t_{L}$, and $\mathcal{I}(t_{L}, t_{k})$ is the information matrix by the measurement at $t_{k}$ and propagated from $t_{k}$ to $t_{L}$. 
The cummulative information obtained by the measurements alone, $\sum_{k=0}^{L-1}\mathcal{I}(t_{L}, t_{k})$ is called the Fisher information matrix\cite{maybeck1982stochastic}. This term is also related to the observability of the discrete system; the system without process noise and measurement noise is observable if and only if the null space of the Fisher information matrix with $R=I$ is $\bm{0}$ for some finite $L$ \cite{maybeck1982stochastic}. For an in-depth exploration of the observability metrics pertinent to cislunar SDA, the reader is referred to the work presented in \cite{fowler2023observability}.

\subsection{Circular Restricted Three-Body Problem}
We model the cislunar dynamics as a circular restricted three-body problem (CR3BP) with the Earth and Moon as the two primaries and the spacecraft as a massless particle. The CR3BP is a simplified model of the three-body problem where the mass of the smaller primary is negligible compared to the larger primary. The CR3BP is a good approximation for the translunar and cislunar regions, where the Moon is the smaller primary and the Earth is the larger primary. 
Let $\bm{r} \in \R^3$ and $\bm{v}\in\R^3$ be the nondimensionalized position and velocity of the spacecraft in the Earth-Moon rotating frame. 
The CR3BP equations of motion are given by
\begin{equation}
\dot{\bm{r}}=\bm{v}, \quad 
\dot{\bm{v}} + 2\bm{\omega}\times\bm{v} = \nabla U(\bm{r})
\end{equation}
where $\bm{\omega}=[0, 0, 1]^T$ and the potential function $U$ is given by
\begin{equation}
\begin{split}
    U &= \frac{1}{2}(x^{2}+y^{2})+\frac{1-\mu}{d_1}+\frac{\mu}{d_2}\\
    d_1^{2}&=(x+\mu)^{2}+y^{2}+z^{2}\\
    d_2^{2}&=(x-1+\mu)^{2}+y^{2}+z^{2}
    \label{eq:potential_cr3bp}
\end{split}
\end{equation}
with $\mu$ representing the mass parameter. $d_1$ and $d_2$ represent the distances from the spacecraft to Earth and Moon, respectively. The state-transition matrix $\Phi(t, t_0)$ is propagated through the Jacobian of the dynamics
\begin{equation}
    \dot{\Phi}(t_{k}, t_0) = 
    \begin{bmatrix}
        \bm{0} & I\\
        \nabla^2 U & \Omega
    \end{bmatrix}\Phi(t_{k}, t_0), \quad
    \Omega = \begin{bmatrix}
        0 & 2 & 0\\
        -2 & 0 & 0\\
        0 & 0 & 0
    \end{bmatrix}.
\end{equation}
Given the state-transition matrix, $\Phi(t_{k}, t_{k-1})$, the state error dynamics is obtained. 
\begin{equation}
    \label{eq:cr3bp-error-dynamics}
    \bm{\xi}(t_{k}) = \Phi(t_{k}, t_{k-1})\bm{\xi}(t_{k-1}), \quad \bm{\xi}(t_{k-1}) = \bm{x}(t_{k-1}) - \bar{\bm{x}}(t_{k-1})
\end{equation}

\begin{table}
    \centering
    \caption{Parameters of CR3BP dynamics}
    \begin{tabular}{p {8cm} p {5cm}}
    \hline
    Parameter & Value\\
    \hline
    Earth-Moon system mass parameter $\mu$, $\mathrm{n.d.}$   & 0.01215058560962404  \\
    Canonical length scale $\mathrm{LU}$, $\mathrm{km}$       & 389703.264829278 \\
    Canonical time scale $\mathrm{TU}$, $\mathrm{s}$          & 382981.289129055  \\
    \hline
    \label{tab:cr3bp}\\
    \end{tabular}
\end{table}

\section{Measurement Model and Uncertainty Deformation}
We consider optical measurements of a target spacecraft from an observer spacecraft, which is applicable to uncooperative targets. The measurements are often defined for the azimuth and elevation angles, but instead we use the directional cosine vector and its rate to exploit the symmetricity of the expression and also to avoid the singularity at zenith and nadir. For an analytical approximations for the Fisher information matrix using the azimuth and elevation, the reader is referred to the work presented in \cite{miga2022analytical}.

Let $\bm{r}$ and $\bm{v}$ be the position and velocity, and the subscript $i$ and $j$ denote the observer and target, respectively. The relative position and velocity vectors are given by
\begin{equation}
    \label{eq:obs-relative}
    \bm{r}_{ij} = \bm{r}_j - \bm{r}_i,\quad \bm{v}_{ij} = \bm{v}_j - \bm{v}_i.
\end{equation}
Our measurements are the directional cosine vector and its rate.
\begin{equation}
\bm{y}_{ij} = \frac{\bm{r}_{ij}}{r_{ij}},\quad \dot{\bm{y}}_{ij} = \frac{\bm{v}_{ij}}{r_{ij}} - \frac{(\bm{r}_{ij}^T\bm{v}_{ij})\bm{r}_{ij}}{r_{ij}^3}, \quad r_{ij}=\|\bm{r}_{ij}\|
\end{equation}

We derive the linearlized measurement error model with respect to the error state of the target, $\bm{\xi}_{j}$. Let $\bm{\zeta}_{ij}=\bm{y}_{ij} - \bar{\bm{y}}_{ij}$ and $\dot{\zeta}_{ij}=\dot{\bm{y}}_{ij} - \bar{\dot{\bm{y}}}_{ij}$ be the measurement error and its rate, respectively. The linearized measurement error model is given by
\begin{equation}
    \label{eq:linear-measurement}
    \begin{bmatrix}
    \bm{\zeta}_{ij}\\
    \dot{\bm{\zeta}}_{ij}
    \end{bmatrix}
    =H_{ij}\bm{\xi}_{j} + \bm{v}
\end{equation}
where $H_{ij}$ is the Jacobian matrix of the measurement model. 
We assume that the measurement noise variance is fixed for the directional cosine vector, and proportional to the exposure time for the directional cosine rate. The measurement noise $\bm{v}$ is given by
\begin{equation}
    \label{eq:obs-noise}
    \bm{v}\sim\sN(\bm{0}, R), \quad
    R = \sigma^2 \begin{bmatrix}
    I & 0\\
    0 & \frac{2}{\Delta t^2}I
    \end{bmatrix}
\end{equation}
where $\sigma$ is the standard deviation of the angle measurement noise and $\Delta t$ is the exposure time.

\subsection{Jacobian Matrix}
The Jacobian matrix $H_{ij}$ is obtained by taking the derivative of the measurement model with respect to the target state; we assume that the observer state is known. Although the measurements $\bm{y}_{ij}$ and $\dot{\bm{y}}_{ij}$ are functions of the relative position and velocity vectors $\bm{r}_{ij}$ and $\bm{v}_{ij}$, due to their relation to the target state by Eq. \eqref{eq:obs-relative}, we can compute the Jacobian matrix $H_{ij}$ by taking the derivative of the measurement model with respect to the target state. The Jacobian matrix is given by
\begin{equation}
    \label{eq:obs-jacobian-ij}
    H_{ij} = \begin{bmatrix}
        \dfrac{\partial\bm{y}_{ij}}{\partial \bm{r}_j}(\bar{\bm{r}}_{ij}) & \bm{0}\\
        \dfrac{\partial\dot{\bm{y}}_{ij}}{\partial \bm{r}_j}(\bar{\bm{r}}_{ij}, \bar{\bm{v}}_{ij}) & \dfrac{\partial\dot{\bm{y}}_{ij}}{\partial \bm{v}_j}(\bar{\bm{r}}_{ij}, \bar{\bm{v}}_{ij})
    \end{bmatrix}
    = \begin{bmatrix}
        H_{ij,11} & \bm{0}\\
        H_{ij,21} & H_{ij,22}
    \end{bmatrix}
\end{equation}
where 
\begin{equation}
    \label{eq:obs-jacobian-ij-11}
    \begin{aligned}
        &H_{ij,11} = H_{ij,22}=
        \frac{I}{\bar{r}_{ij}}-\frac{\bar{\bm{r}}_{ij}\bar{\bm{r}}_{ij}^T}{\bar{r}_{ij}^3}\\
        &H_{ij,21} = 
        -\frac{\bar{\bm{v}}_{ij}\bar{\bm{r}}_{ij}^T}{\bar{r}_{ij}^3} 
        -\frac{\bar{\bm{r}}_{ij}\bar{\bm{v}}_{ij}^T + \bar{\bm{r}}_{ij}^T\bar{\bm{v}}_{ij}I}{\bar{r}_{ij}^3} + 3\frac{(\bar{\bm{r}}_{ij}^T\bar{\bm{v}}_{ij})\bar{\bm{r}}_{ij}\bar{\bm{r}}_{ij}^T}{\bar{r}_{ij}^5}.\\
    \end{aligned}
\end{equation}

\subsection{Null Space}
The angle and angular rate measurements do not provide the range and range rate information, and the measurement Jacobian has rank of 4 at maximum. We can confirm this by analytically identifying the null space of the measurement Jacobian. The null space of the Jacobian matrix $H_{ij}$ is given by
\begin{equation}
    \label{eq:obs-null}
    \begin{aligned}
        \text{Null}(H_{ij}) &= \left\{\bm{v}\in\R^6\mid H_{ij}\bm{v}=\bm{0}\right\}\\
        &= \left\{\bm{v}\in\R^6\mid \bm{v} = \alpha \bm{v}_1 + \beta \bm{v}_2, \quad \alpha, \beta\in\R\right\}
    \end{aligned}
\end{equation}
where
\begin{equation}
    \label{eq:obs-null-v1v2}
    \bm{v}_1 = \begin{bmatrix}
        \bar{\bm{r}}_{ij}\\
        \bar{\bm{v}}_{ij}
    \end{bmatrix}
    ,\quad
    \bm{v}_2 = \begin{bmatrix}
        \bm{0}\\
        \bar{\bm{r}}_{ij}
    \end{bmatrix}.
\end{equation}
The null space of the Jacobian matrix $H_{ij}$ is spanned by the vectors $\bm{v}_1$ and $\bm{v}_2$. The direction of $\bm{v}_1$ corresponds to the relative position and the velocity, which is because the measurement model is defined only with the relative position and velocity directions of the target and the observer, and the range and range rate information cannot be observed. The direction of $\bm{v}_2$ is the case when the relative velocity change is in the direction of the relative position, which also cannot be detected by the angle and angular rate measurement. We can analytically check the null space. 
\begin{equation}
    \begin{aligned}
        &H_{ij}\bm{v}_1 = \begin{bmatrix}
            H_{ij,11}\bar{\bm{r}}_{ij}\\
            H_{ij,21}\bar{\bm{r}}_{ij} + H_{ij,22}\bar{\bm{v}}_{ij}
        \end{bmatrix}
        = \begin{bmatrix}
            \bm{0}\\
            \left(-\dfrac{\bar{\bm{v}}_{ij}}{\bar{r}_{ij}}+\dfrac{(\bar{\bm{v}}_{ij}^T\bar{\bm{r}}_{ij})\bar{\bm{r}}_{ij}}{\bar{r}_{ij}^3}\right)+\left(\dfrac{\bar{\bm{v}}_{ij}}{\bar{r}_{ij}}-\dfrac{(\bar{\bm{v}}_{ij}^T\bar{\bm{r}}_{ij})\bar{\bm{r}}_{ij}}{\bar{r}_{ij}^3}\right)
        \end{bmatrix}
        = \begin{bmatrix}
            \bm{0}\\
            \bm{0}
        \end{bmatrix}\\
        &H_{ij}\bm{v}_2 = \begin{bmatrix}
            \bm{0}\\
            H_{ij,22}\bar{\bm{r}}_{ij}
        \end{bmatrix} 
        = \begin{bmatrix}
            \bm{0}\\
            \bm{0}
        \end{bmatrix}
    \end{aligned}
\end{equation}

\subsection{Informative Space}
The information gain by the measurement is given by $H^T R^{-1} H$, as shown in Eq. \eqref{eq:eif-update}. The null space of the information gain is given by $\text{Null}(H^T R^{-1} H)=\text{Null}(H)$, and the informative space is spanned by four linearly independent eigenvectors, which are orthogonal to the null space. For simplicity, we omit the subscript $ij$ in the following discussion. The information gain is given by
\begin{equation}
    \label{eq:obs-info-gain}
    H^T R^{-1} H = \sigma^{-2}\begin{bmatrix}
        H_{11}^TH_{11} + H_{21}^TH_{21} & \dfrac{\Delta t^2}{2}H_{21}^TH_{11}\\
        \dfrac{\Delta t^2}{2}H_{22}^TH_{21} & \dfrac{\Delta t^4}{4}H_{22}^TH_{22}
    \end{bmatrix}.
\end{equation}
Numerically we identified that the informative space often has two major eigenvalues and two minor eigenvalues, resulting in a "thin" four dimensional ellipsoid, which is qualitatively similar to a two dimensional plane. Informally, we can observe this structure by Eq. \eqref{eq:obs-info-gain} as $\Delta t^2 << 1$; the typical sensor exposure time, $\Delta t$, is about the order of $1e2\sim1e3$ seconds, but the time unit of the cislunar dynamics used for normalization is about $TU=3.8e5$ seconds. Therefore, the eigenspace closer to the upper row of the information matrix is dominant. As our null space spans both in the position and the velocity space, we have two major information space closer to the position space, and two minor information space closer to the velocity space. For example, let $\bm{r}_{\perp,1}\in\R^3$ and $\bm{r}_{\perp,2}\in\R^3$ be the two vectors orthogonal to the relative position vector $\bm{r}_{ij}$. Then, the vectors $[\bm{r}_{\perp, 1}^T, \bm{0}^T]^T$ and $[\bm{r}_{\perp, 2}^T, \bm{0}^T]^T$ both have the same eigenvalue of $\sigma^{-2}/r_{ij}^2$. On the other hand, the vector $[\bm{0}^T, \bm{rv}_{\perp}^T]^T$ where $\bm{rv}_{\perp}\in\R^3$ is a vector orthogonal to $\bm{r}_{ij}$ and $\bm{v}_{ij}$, has the eigenvalue of $\sigma^{-2}\Delta t^4/4r_{ij}^2 << \sigma^{-2}/r_{ij}^2$.

\subsection{Uncertainty Deformation}
The measurement information by the allocation of an observer $i$ to a target $j$ at time $k$ is represented by $H_{ijk}^TR^{-1}H_{ijk}$. Following the update equation of the EIF, the information gain propagated to time $t_L$ is obtained as follows:
\begin{equation}
    \mathcal{I}(t_{L}, t_{k}) = \Phi(t_{k}, t_{L})^T H(t_{k})^T R^{-1} H(t_{k})\Phi(t_{k}, t_{L})
\end{equation}
where $\Phi(t_{k}, t_{L})$ is the backward state transition matrix from $t_L$ to $t_k$. We provide the following theorem to highlight the propagated information's physical meanings. 

\begin{theorem} \label{theorem1}
    Let $\sigma_{\text{max}}(\cdot)$ and $\sigma_i(\cdot)$ denote the maximum singular value and the $i$-th largest singular value of an input matrix, respectively. Suppose $R$ be a symmetric positive definite matrix. Then, $\sigma_5(H^TR^{-1}H)=\sigma_6(H^TR^{-1}H)=0$ and we have the following bounds:
    \begin{equation}\label{eq:bounds}
        \sigma_{\text{max}}(\Phi\Phi^T)\sum_{i=1}^4 \alpha_i^2\sigma_i(H^TR^{-1}H) \leq \sigma_{\text{max}}(\Phi^TH^TR^{-1}H\Phi) \leq  \sigma_{\text{max}}(\Phi\Phi^T)\sigma_{\text{max}}(H^TR^{-1}H)
    \end{equation}
    where $\alpha_i$ represents the alignment between the $i$-th largest eigenvector of $H^T R^{-1} H$, denoted by $\bm{v}_{i}$, and the largest eigenvector of the left Cauchy-Green tensor, $\Phi\Phi^T$, denoted by $\bm{v}_{\text{CGT}}$:
    \begin{equation}
    \label{eq:alignment}
        \alpha_i = \langle \bm{v}_{i} \; , \; \bm{v}_{\text{CGT}} \rangle, \quad i=1, 2, 3, 4.
    \end{equation}
\end{theorem}

Proof: See Appendix A. 

Theorem \ref{theorem1} implies that the propagated information $\mathcal{I}(t_{L}, t_{k})$ is maximized when the measurement information $H^TR^{-1}H$ and the uncertainty deformation in the form of the left Cauchy-Green tensor, $\Phi\Phi^T$, are not only maximized but also aligned well. In other words, the information is maximized when we have a good state measurement in the direction to which the uncertainty is extended. This is a key insight for Cislunar SDA where the spacecraft state uncertainty undergoes large deformation due to the chaotic nature of the dynamics. In this regard, predictive sensor tasking is advantageous as it can utilize the prospective uncertainty deformation by the dynamics to maximize the observed information.

\section{Multi-Observer, Multi-Target Predictive Sensor Tasking}
\subsection{Design Variables}
\begin{figure}[htbp]
    \centering
    \includegraphics[width=0.5\textwidth]{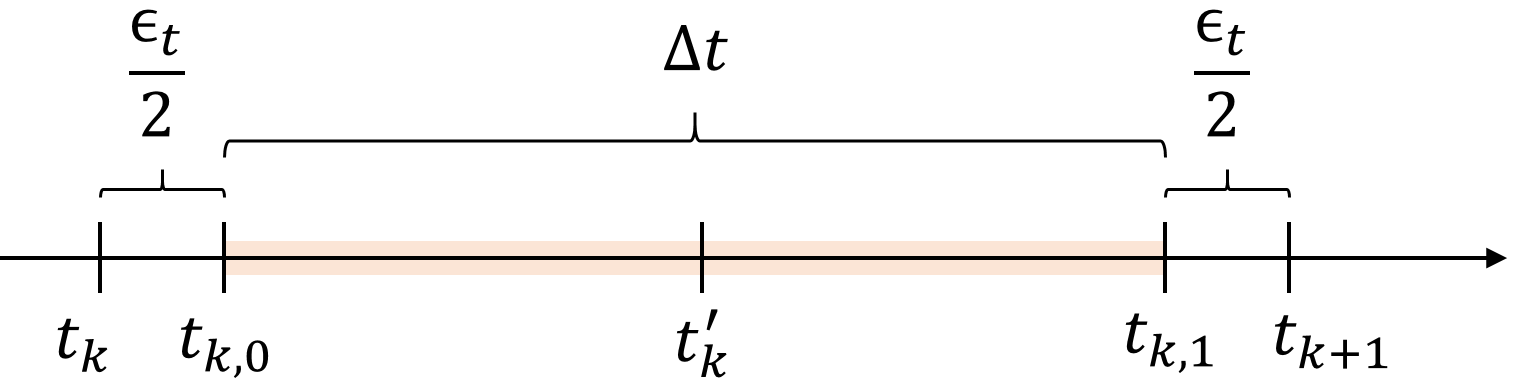}
    \caption{Time steps for the predictive sensor tasking where $t_k$ and $t_{k+1}$ are the time instances for decision making. The sensor exposure starts at $t_{k,0}$ and ends at $t_{k,1}$.
    We approximate the optical measurement obtained by the exposure time of $\Delta t$ by the instantaneous measurement at $t_{k}'=(t_{k,0} + t_{k,1})/2$. Here $\Delta t$ is the exposure time and $\epsilon_t$ is the buffer for sensor steering.}
    \label{fig:time-step}
\end{figure}

We consider the predictive sensor tasking problem with time horizon $L$ for the fixed set of time instances $\{t_k\}_k =\{t_0, ..., t_{L-1}\}$. Optical observation requires finite exposure time, $\Delta t$, and the wait time between observations including the time for steering, $\epsilon_t$.
Figure \ref{fig:time-step} shows the time step for the predictive sensor tasking. 
The time instances for decision making satisfy $t_{k+1}= t_k + \Delta t + \epsilon_t$ for $k=0, ..., L-1$. The sensor exposure starts at $t_{k,0}$ and ends at $t_{k,1}$.
We approximate the optical measurement obtained by the exposure time of $\Delta t$ by the instantaneous measurement at $t_{k}'=(t_{k,0} + t_{k,1})/2$. 

We optimize the sensor allocations at the decision making time instances $\{t_k\}_k =\{t_0, ..., t_{L-1}\}$. Our design variables are the binary tasking allocation variables $u_{ijk}$ where $i=1,\cdots,M$, $j = 1,\cdots,N$, and $k=0,\cdots,L-1$. Here $M$, $N$, $L$ denote the number of the observers, targets, and time steps, respectively. The binary variable $u_{ijk}$ represents the tasking allocation of the $i$th observer to the $j$th space object at time step $k$; if $u_{ijk}=1$, the $i$th observer is tasked to the $j$th space object at time step $k$. Otherwise, $u_{ijk}=0$.
The tasking allocation variables $u_{ijk}$ are subject to the following constraints:
\begin{equation}
    \label{eq:single-observation}
    \sum_{j=1}^{N} u_{ijk} \leq 1, \quad \forall i, k
\end{equation}
\begin{equation}
    \label{eq:binary-cstr}
    u_{ijk} \in \{0, 1\}, \quad \forall i, j, k.
\end{equation}
Eq. \eqref{eq:single-observation} ensures that each observer is tasked to at most one target space object at each time and Eq. \eqref{eq:binary-cstr} states that the tasking allocation is binary. 
Since we need at least two independent measurements to make the state observable, we also add a constraint that all objects are observed at least twice during the time horizon:
\begin{equation}
    \label{eq:all-observed}
    \sum_{i=1}^M \sum_{k=0}^{L-1} u_{ijk} \geq 2, \quad \forall j.
\end{equation}

\subsection{Objective Function}
Our main objective is to minimize the state uncertainty of all space objects at the end of the time horizon. To maintain the linear property of the objective function in terms of the design variables, we use the information form of the covariance matrix. The information matrix $\Lambda$ is the inverse of the covariance matrix $P$ and since $P$ is symmetric positive definite, the following equality holds:
\begin{equation}
    \label{eq:I-P}
    \text{tr}(\Lambda) = \sum_{m=1}^6 \lambda_{m}^{-1}, \quad \text{tr}(P) = \sum_{m=1}^6 \lambda_{m}
\end{equation}
where $\lambda_{m}$ is the $m$th eigenvalue of $P$. 
Note that restricting linear operations of the information form does not allow us to bound the minimum eigenvalue of the information matrix, which corresponds to the maximum eigenvalue of the covariance matrix. However, we empirically show that the objective functions based on the information matrix is a good surrogate for the minimization of the state uncertainty.
\subsubsection{Maximizing Total Information}
Since we do not consider the uncertainty in the observer states, the information matrix are independent between the targets. Let $H_{ijk}$ be the measurement Jacobian of Eq. \eqref{eq:obs-jacobian-ij} evaluated at $t_k'$. Then, maximization of the total observed information at $t_L$ becomes the sum of cummulative information of each spface object as follows:
\begin{equation}
    \label{eq:objective-max-trace}
    \max_{u_{ijk}} \sum_{i=1}^{M}\sum_{j=1}^{N}\sum_{k=0}^{L-1} u_{ijk} \text{tr}\left(\mathcal{I}_{ij}(t_L, t_k')\right), \quad 
    \mathcal{I}_{ij}(t_L, t_k') = \Phi_j(t_k', t_L)^T H_{ijk}^T R^{-1} H_{ijk} \Phi_j(t_k', t_L)
\end{equation}
where $\mathcal{I}_{ij}(t_L, t_k')$ is the information matrix of the $j$th space object due to the measurement of the $i$th observer at time $t_k'$, projected to the time $t_L$. 
\subsubsection{Maximizing Minimum Target Information}
We often want to maximize the minimum cummulative information for the space objects in the system, instead of the total information. In such cases, we can employ the max-min formulation.
\begin{equation}
    \label{eq:objective-max-min}
    \max_{u_{ijk}} \min_{j}\sum_{i=1}^{M}\sum_{k=0}^{L-1} u_{ijk} \text{tr}\left(\mathcal{I}_{ij}(t_L, t_k')\right).
\end{equation}
Note that this formulation makes the problem mixed-integer linear programming, becase the lower bound of the trace is a continuous variable, which is maximized.

\section{Numerical Analysis}

\label{sec:orbits}
\begin{figure}[htbp]
    \centering
    \begin{subfigure}{0.45\textwidth}
        \centering
        \includegraphics[width=\linewidth]{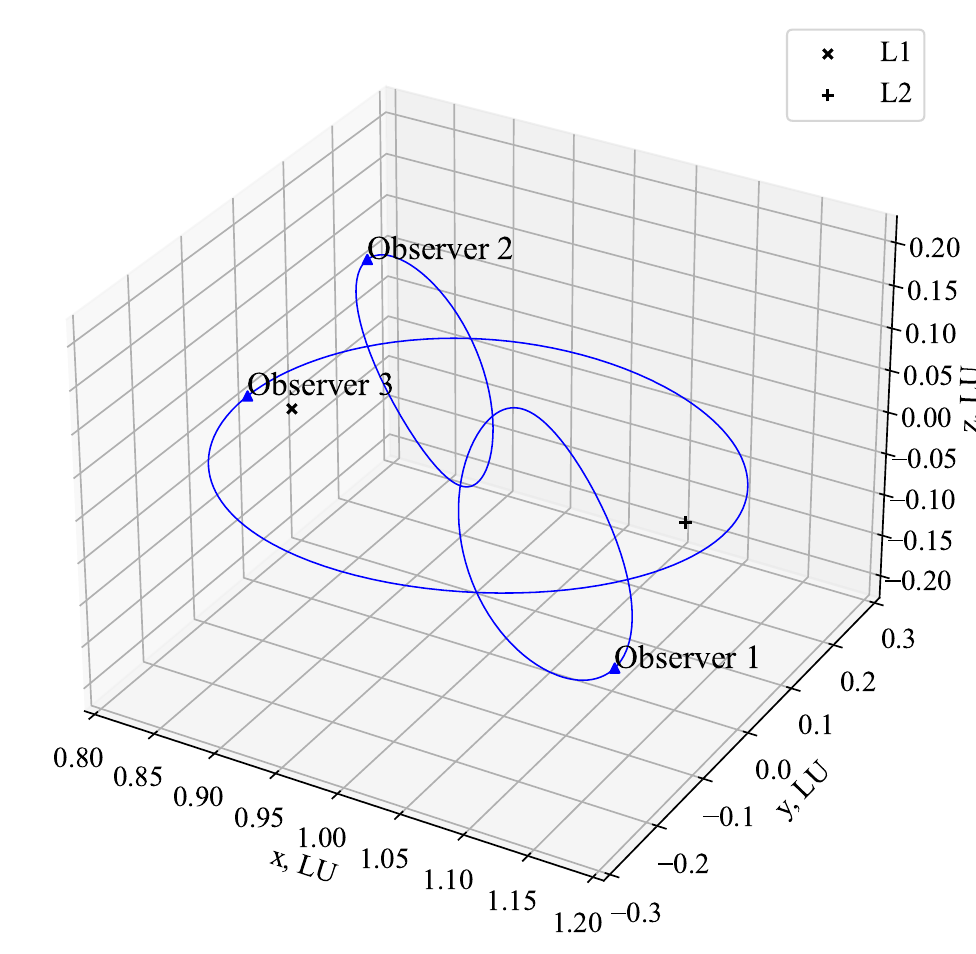}
        \caption{Observer LPOs.}
        \label{fig:sub1}
    \end{subfigure}
    \hfill
    \begin{subfigure}{0.45\textwidth}
        \centering
        \includegraphics[width=\linewidth]{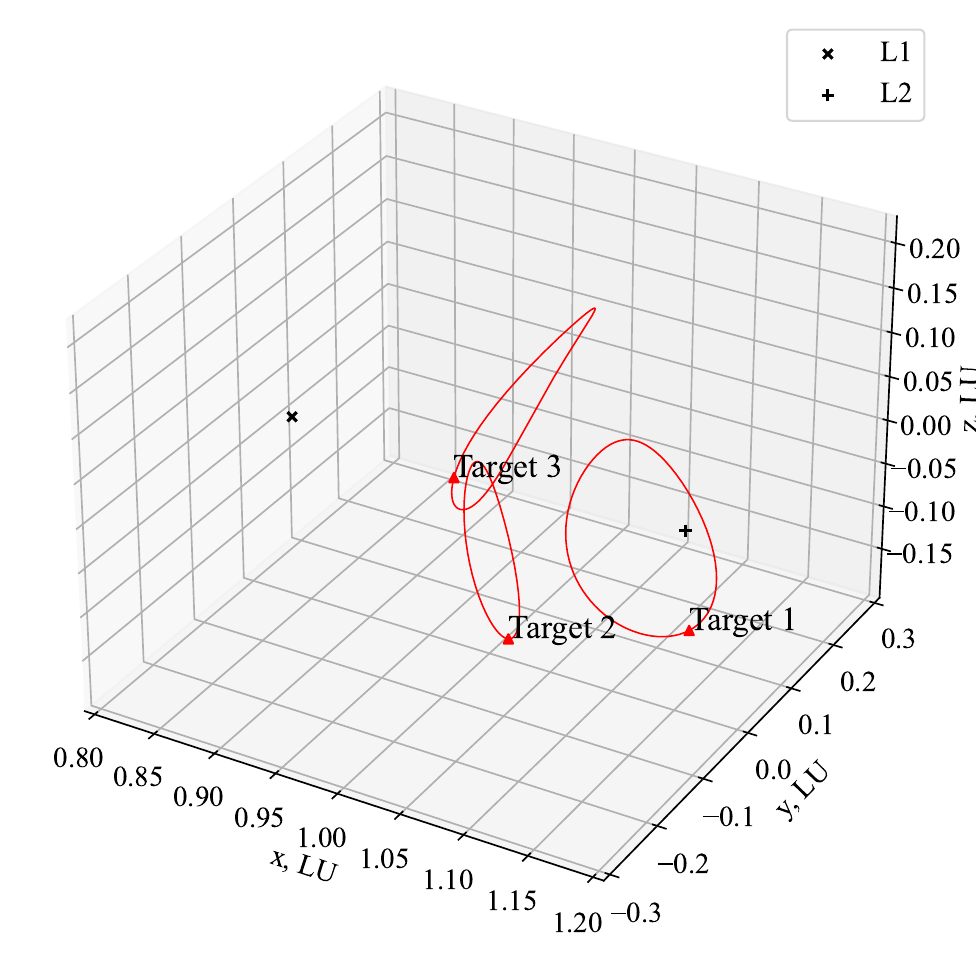}
        \caption{Target LPOs in L2 family.}
        \label{fig:sub2}
    \end{subfigure}
    \vspace{1em}
    \begin{subfigure}{0.45\textwidth}
        \centering
        \includegraphics[width=\linewidth]{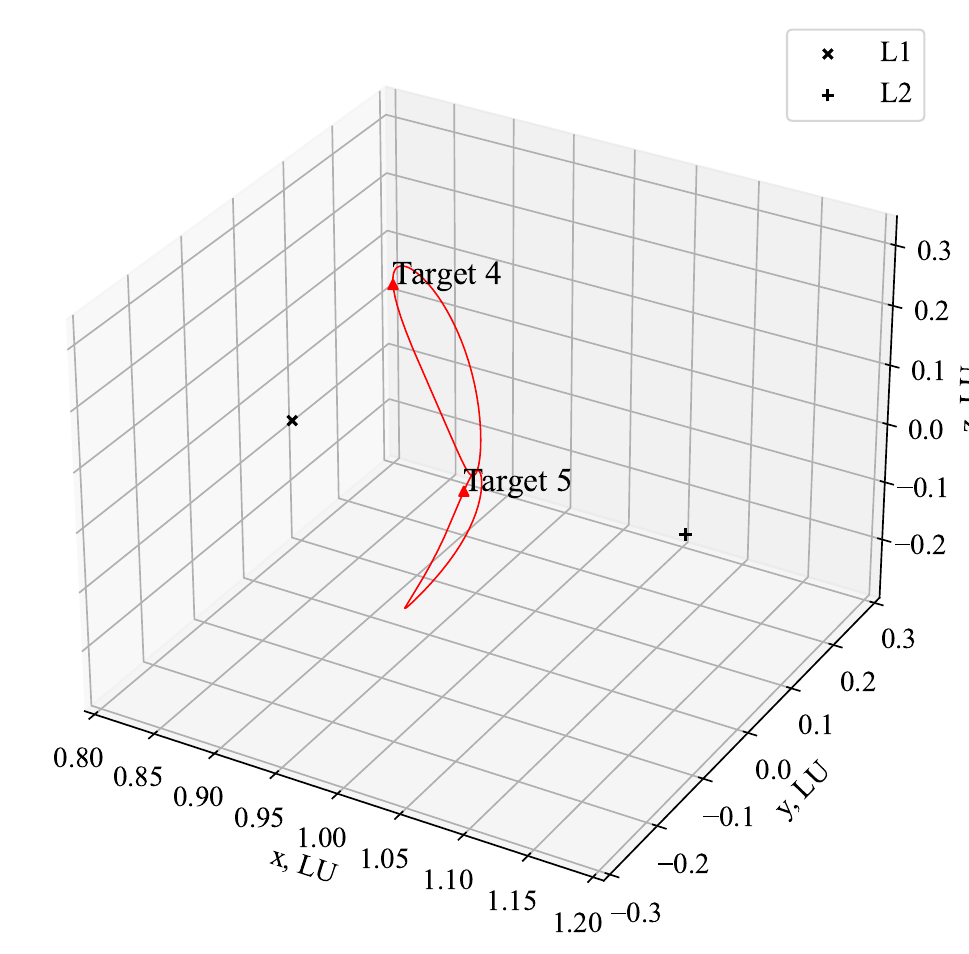}
        \caption{Target LPOs in L1 family}
        \label{fig:sub3}
    \end{subfigure}
    \hfill
    \begin{subfigure}{0.45\textwidth}
        \centering
        \includegraphics[width=\linewidth]{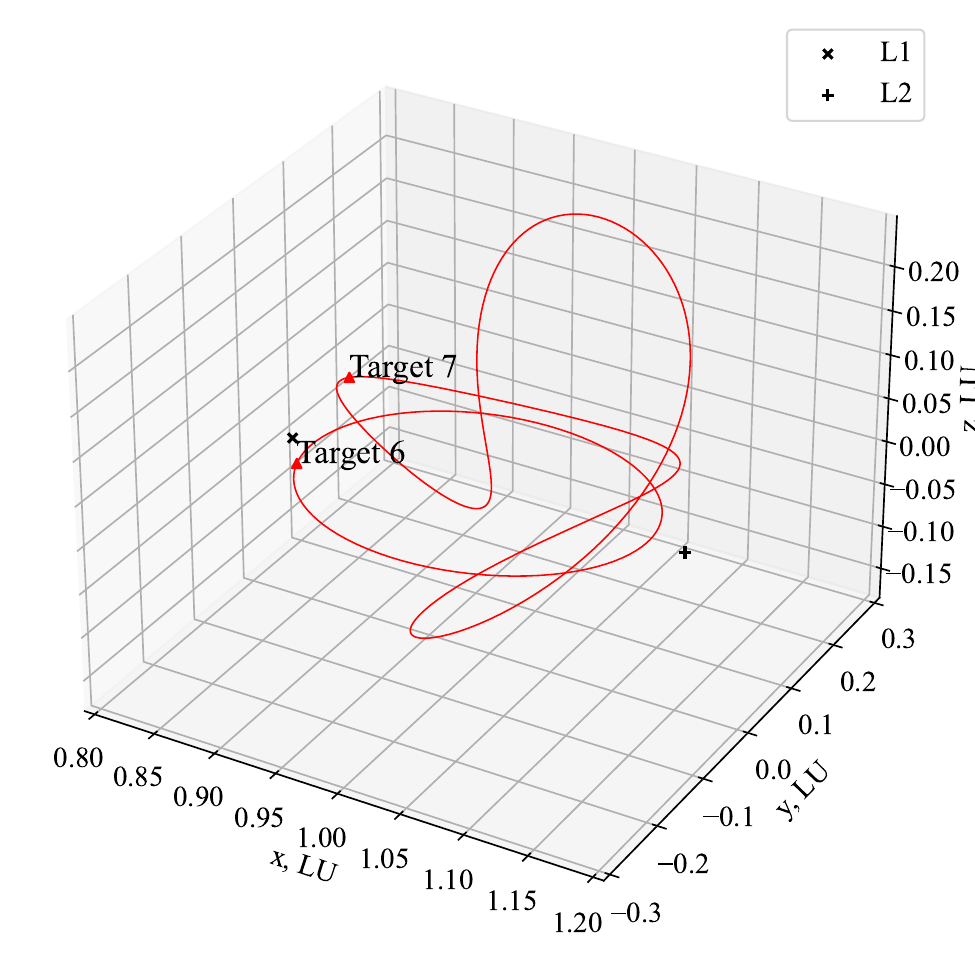}
        \caption{Target LPOs of DRO and Dragonfly orbits.}
        \label{fig:sub4}
    \end{subfigure}
    \caption{Orbits of the observers and the targets. The observer orbits are shown in blue and the target orbits are shown in red.}
    \label{fig:observer-target-orbits}
\end{figure}

We consider 3 observers and 7 targets in the Earth-Moon libration point orbits. The parameters for the observer orbits and the target orbits are shown in Table \ref{tab:observers} and \ref{tab:targets}, respectively. 
Figure \ref{fig:observer-target-orbits} shows the orbits of the observers and the targets.

\begin{table}
    \centering
    \caption{Observer Libration Point Orbits}
    \begin{tabular}{p {0.5cm} p {3.2cm} p {1.5cm} p {1.5cm} p {1.5cm} p {1.5cm}}
    \hline
    ID & Libration Point Orbit & Period, TU & Synodic resonance & Stability index & 
    Phase
    \\
    \hline
    1 & L2 Halo Southern  & 2.66 & 5:2 & 7.00 & 0.00\\
    2 & L1 Halo Northern  & 1.90  & 7:2 & 2.08 & 0.00\\
    3 & DRO               & 3.33  & 2:1 & 1.00 & 0.00\\
    \hline
    \label{tab:observers}\\
    \end{tabular}
\end{table}

\begin{table}
    \centering
    \caption{Target Libration Point Orbits}
    \begin{tabular}{p {0.5cm} p {3.2cm} p {1.5cm} p {1.5cm} p {1.5cm} p {1.5cm}}
    \hline
    ID & Libration Point Orbit & Period, TU & Synodic resonance & Stability index & 
    Phase
    \\
    \hline
    1 & L2 Halo Southern  & 3.33 & 2:1 & 2.91e2 & 3.38e-2\\
    2 & L2 Halo Southern  & 1.48  & 9:2 & 1.26 & 6.45e-2\\
    3 & L2 Halo Northern   & 2.22  & 3:1 & 1.00 & 4.03e-1\\
    4 & L1 Halo Northern   & 2.22 & 3:1 & 2.14 & 8.91e-1\\
    5 & L1 Halo Southern   & 2.00  & 10:3 & 2.74 & 5.11e-1\\
    6 & DRO Northern   & 2.22  & 3:1 & 1.00 & 9.57e-1\\
    7 & Dragonfly Northern   & 5.55  & 1:1 & 2.25e2 & 1.92e-1\\
    \hline
    \label{tab:targets}\\
    \end{tabular}
\end{table}

\subsection{Predictive Tasking Performance}
We compared the predictive algorithms for maximizing the total trace of the system and maximizing the minimum trace of the targets, corresponding to Eqs. \eqref{eq:objective-max-trace} and \eqref{eq:objective-max-min}, respectively. The predictive algorithms are compared with the memory-less myopic policy; the myopic policy maximizes the total trace of the information gain each time. Table \ref{tab:objective-comparison} shows the comparison of the performance of the tasking algorithms. 
\begin{table}
    \centering
    \caption{Comparison of the tasking algorithm performance.}
    \begin{tabular}{p{3.2cm} p{2.5cm} p{2.8cm} p{2.8cm} p{2.8cm}}
    \hline
    Algorithms & $\sum_{j}\text{Tr}(\mathcal{I}_j(t_f))$ & $\min_{j}\text{Tr}(\mathcal{I}_j(t_f))$ & $ \max_j\sigma_{\text{max}}(\mathcal{I}_j(t_f))$ & $\min_j\sigma_{\text{max}}(\mathcal{I}_j(t_f))$\\
    \hline
    Myopic, Max               & 1.846e+13  & 4.867e+09 & 7.955e+12 & 1.035e+08\\
    Predictive, Max     & \textbf{2.807e+13}  & 6.868e+09 & \textbf{9.575e+12} & 1.035e+08\\
    Predictive, MaxMin  & 3.885e+12  & \textbf{4.416e+11} & 1.004e+12 & \textbf{1.036e+08}\\
    \hline
    \label{tab:objective-comparison}\\
    \end{tabular}
\end{table}
The max-trace predictive algorithm with the objective function of Eq. \eqref{eq:objective-max-trace} outperforms the myopic policy in terms of the total information gain, $\sum_{j}\text{Tr}(\mathcal{I}_j)$ and the maximum singular value of the system, $ \max\sigma_{\text{max}}(\mathcal{I}_j)$, without sacrificing the other metrics. The min-max-trace predictive algorithm with the objective function of Eq. \eqref{eq:objective-max-min} outperforms the myopic policy in terms of the minimum trace of the information gain, $\max\min_{j}\text{Tr}(\mathcal{I}_j)$, without sacrificing the other metrics.

\begin{figure}[htbp]
    \centering
    \includegraphics[width=0.99\textwidth]{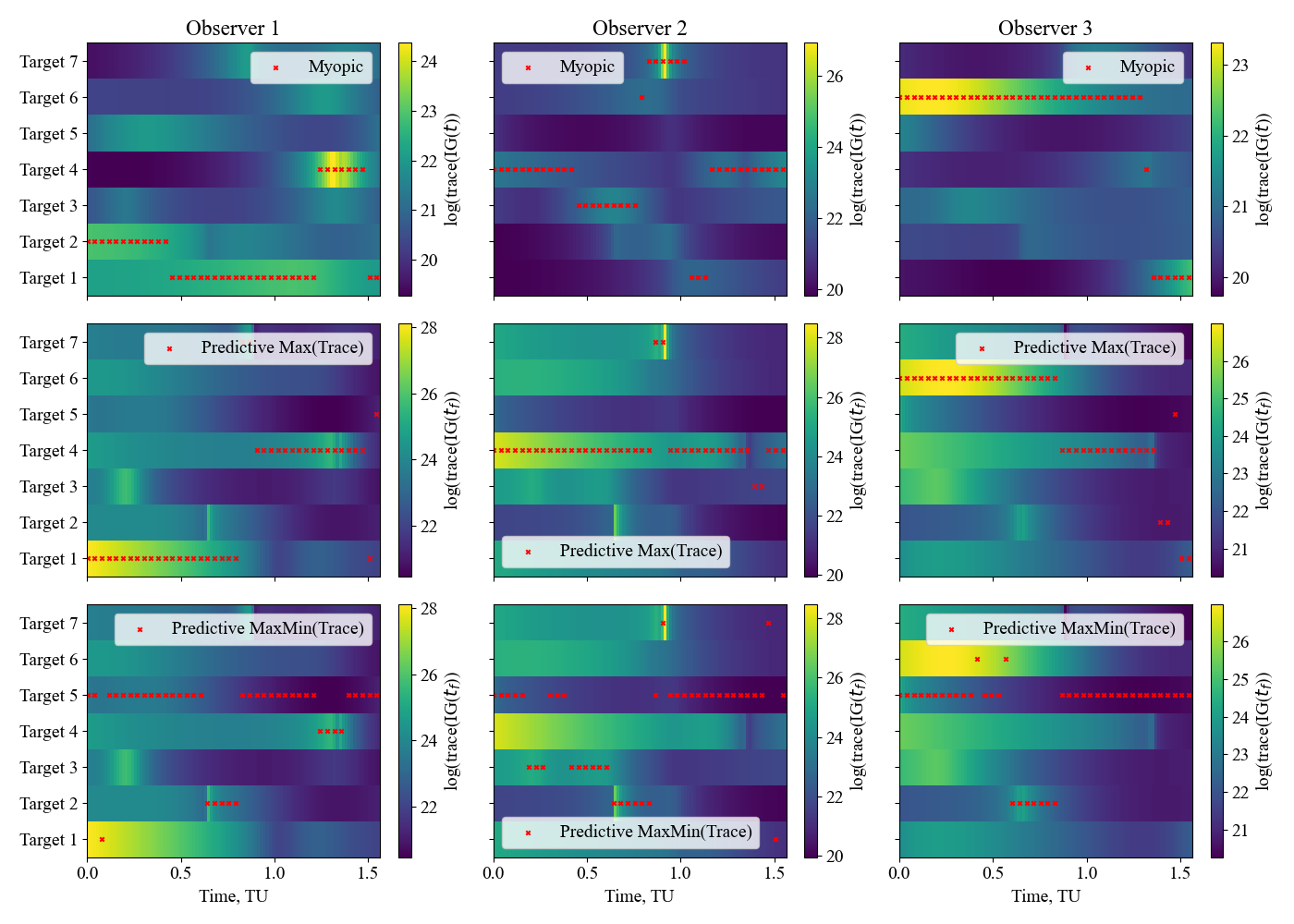}
    \caption{Comparison of the observer-target allocations with the measurement information, at the time of measurement for the myopic algorithm, and propagated to the final time for the predictive algorithms.}
    \label{fig:actions}
\end{figure}

Figure \ref{fig:actions} compares the observer-target allocations with the measurement information. For the myopic policy, the color shows the amout of information at the time of measurement, whereas it shows the projected information to the final time, $t_L$, for the two predictive algorithms. First, we can see that there exists a clear difference in the trend between the information evaluated at the observation time and the information evaluated at a fixed later time, $t_L$.
Next, the myopic policy and max-trace predictive algorithm both tend to allocate the observers to the targets with the largest information gain, which is projected to the final time, $t_L$, for the predictive algorithm. On the other hand, the min-max-trace predictive algorithm tends to allocate the observers to the targets with the smallest information gain. 

\subsection{Deformed Information}
\begin{figure}[htbp]
    \centering
    \includegraphics[width=0.9\textwidth]{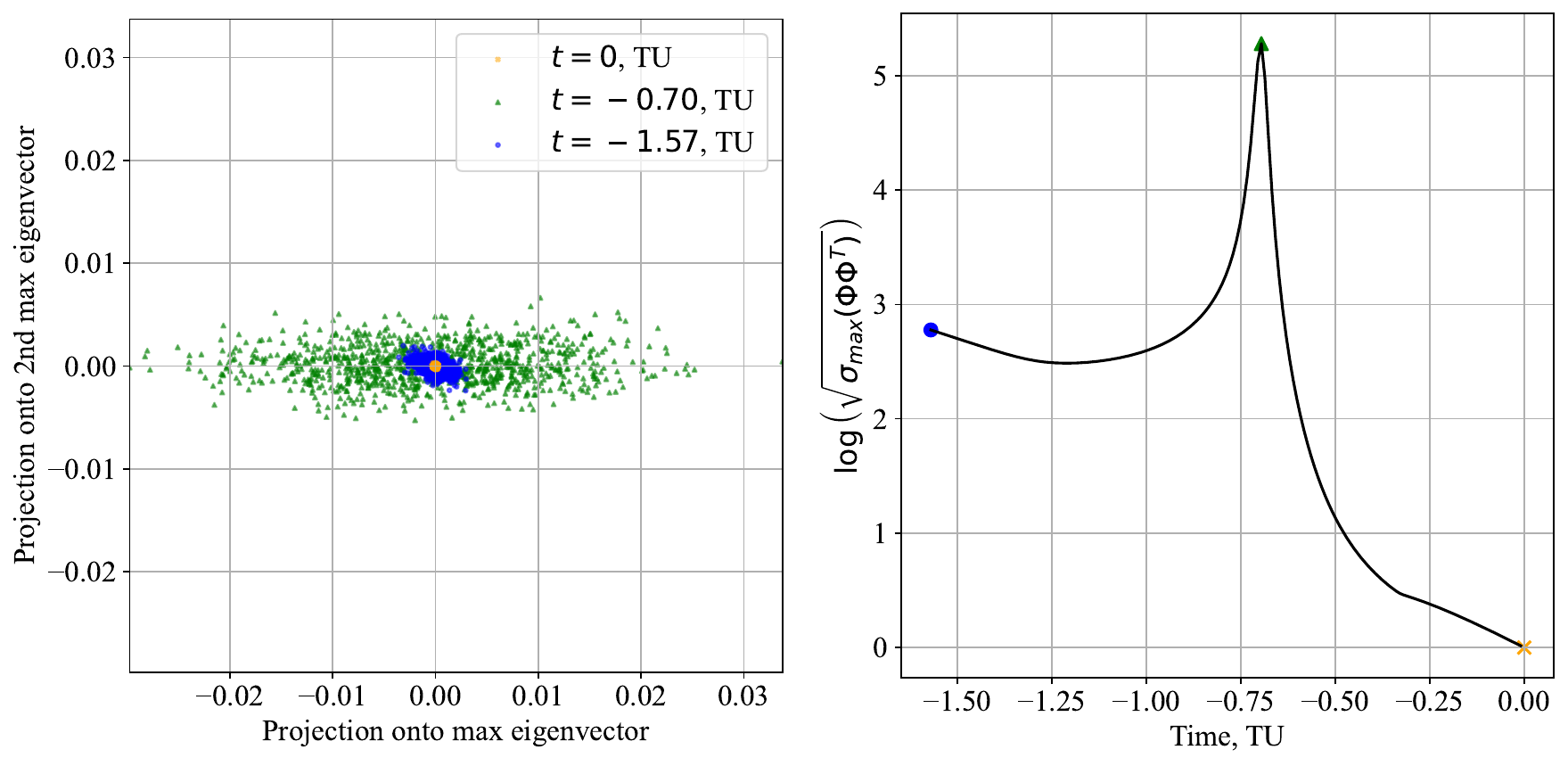}
    \caption{The right plot shows the backward transition of the magnitude of uncertainty deformation from the reference time of $t_L$, where $\sigma_{\text{max}}(\Phi\Phi^T)$ is the maximum singular value of the left CGT. The left plot is the error distributions projected onto the first and second largest eigenvectors of the left CGT.}
    \label{fig:deformed-info}
\end{figure}

Next, we look into the effect of the cislunar dynamics on the measurement information that is propagated to the evaluation time. 
Figure \ref{fig:deformed-info} demonstrates that uncertainty deformation can be non-monotonic with respect to time for Cislunar LPOs, showing the case for target 7. The right plot of Figure \ref{fig:deformed-info} shows the magnitude of uncertainty deformation from the evaluation time $t_L$ backward, where $\sigma_{\text{max}}(\Phi\Phi^T)$ is the maximum singular value of the left CGT. Starting from the reference time $t=0$, the deformation reaches its peak at $t=-7.0$ TU, and shrinks toward $t=-1.57$ TU. The left plot of Figure \ref{fig:deformed-info} validates this non-monotonic trend of deformation with the error distributions of randomly perturbed samples. We sampled $1000$ perturbed states at the reference time and propagated them backward, and the left plot of Figure \ref{fig:deformed-info} shows their distributions projected onto the first and second largest eigenvectors of the left CGT, $\Phi\Phi^T$. We can also confirm that the eigenvector of the left CGT, which is equivalent to the left eigenvector of the state transition matrix $\Phi$ from the reference time, represents the directions with the largest deformations. It validates the usage of the left CGT as our interests are the largest deformation directions in the state space at the time after backward propagation from the reference time, instead of at the reference time. If we would like to know the directions that undergo the largest deformation in the state space at the reference time, then the right (=regular) eigenvectors have this information.

\begin{figure}[htbp]
    \centering
    \includegraphics[width=0.99\textwidth]{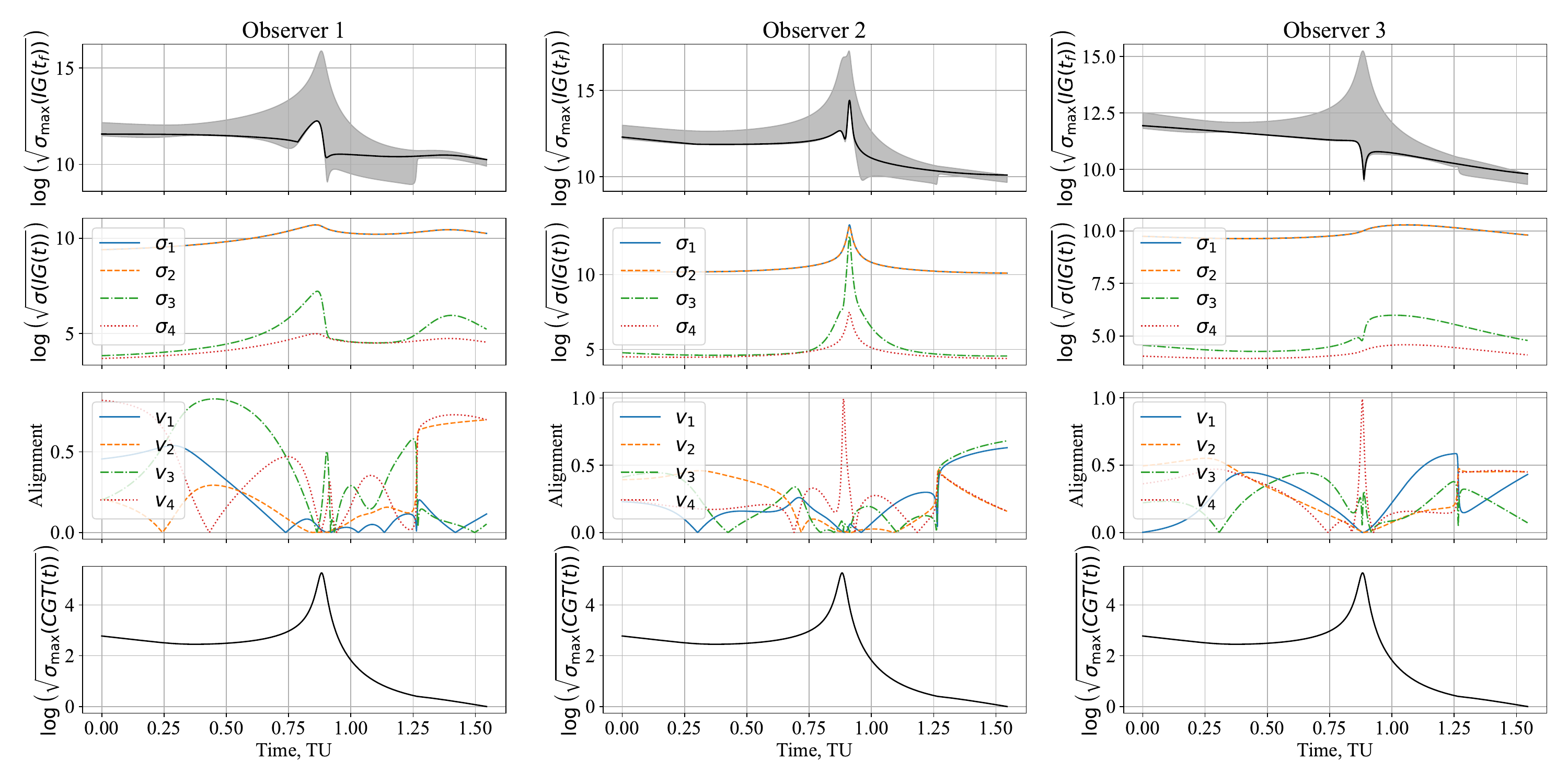}
    \caption{The first row shows the representative magnitude of the propagated information with the derived bounds of Eq. \eqref{eq:bounds}, the second row is the measurement information, the third row is the alignment between the propagated information and the uncertainty deformation, and the last row shows the uncertainty deformation all along time. The columns correspond to each observer. This is for the target 7.}
    \label{fig:cgt-ig-comparison}
\end{figure}

Figure \ref{fig:cgt-ig-comparison} shows that the information propagated to the reference time is maximized when both the measurement information and the uncertainty deformation from the reference time are large, and their information space is aligned, by the example results for target 7.  The first row shows the representative magnitude of the propagated information with the derived bounds of Eq. \eqref{eq:bounds}, the second row is the measurement information, the third row is the alignment between the propagated information and the uncertainty deformation, and the last row shows the uncertainty deformation all along time. The columns correspond to each observer. 
The first row numerically validates the derived bounds of Eq. \eqref{eq:bounds}, and shows that the lower bound is relatively tight. Comparing the four rows, we can observe that the information propagated to the reference time is maximized when both the measurement information and the uncertainty deformation from the reference time are large, and their information space is aligned; for example, the peak of the propagated information for observer 1 and 2 is observed at around $t=0.8$ TU where the peak of the uncertainty deformation exists. However, observer 3 does not have the peak at around $t=0.8$, and this is because the information alignment is weak around that time except for the fourth singular value, which is small. 

\section{Conclusion}
We have investigated the predictive sensor tasking algorithm based on the Extended information filter (EIF). The additive structure of the measurement information of the EIF has been exploited to formulate the linear integer programming formulation for the multi observer, multi target sensor allocation problem. 

The measurement Jacobian is derived with the relative position and velocity of the target and the observer, and their null space and the remaining informative 4D space were investigated. Specifically, the predictive EIF formulation propagates the information gain by the measurement to a reference time, and the propagation is related to the information space deformed by the left Cauchy-Green tensor. 

We numerically showed that the cislunar dynamics expands and shrinks the state uncertainty information along the Libration point orbits, and the projected information gain is maximized when the deformation by the left CGT is aligned with the informative measurement space. The predictive sensor tasking algorithm was compared with the myopic algorithm and demonstrated to outperform the myopic algorithms in terms of the system's total information gain and the minimum information gains over the targets, depending on the formulation. 

\section*{Appendix: Proof of Theorem \ref{theorem1}} \label{proof1}
Since $R$ is a positive definite matrix, we have $\text{nullity}(H^TR^{-1}H)=\text{nullity}(H)=2$, thus $\sigma_5(H^TR^{-1}H)=\sigma_6(H^TR^{-1}H)=0$. 

To obtain the bounds, we use the following fact: for a real square matrix $A\in\R^{n\times n}$, we have
\begin{equation}\label{eq:norm-equality}
    \sigma_{\text{max}}\left(A^T A\right)
    =\sigma_{\text{max}}^2\left(A\right)
    =\|A\|_2^2 
    =\left(\sup_{\|\bm{x}\|\neq\bm{0}}\frac{\|A\bm{x}\|_2}{\|\bm{x}\|_2}\right)^2.
\end{equation}
Since $R$ is symmetric and positive definite, we can define $R^{-1/2}$ that satisfies $R^{-1}=(R^{-1/2})^TR^{-1/2}$. Let $R^{-\frac{1}{2}}H\triangleq S$, then we have
\begin{equation}
    \begin{split}
        \sigma_{\text{max}}(\Phi^TH^TR^{-1}H\Phi)&=\sigma_{\text{max}}^2\left(S\Phi\right)\\
        &\leq \sigma_{\text{max}}^2(S)\sigma_{\text{max}}^2(\Phi)
        =\sigma_{\text{max}}(H^TR^{-1}H)\sigma_{\text{max}}(\Phi\Phi^T).
    \end{split}
\end{equation}
For the lower bound, we apply the singular value decomposition (SVD); the SVD of a real square matrix $A\in\R^{n\times n}$ is represented by $U D V^T$ where $U$ and $V$ are orthogonal matrices and $D$ is a diagonal matrix with singular values $\sigma_1(A), ..., \sigma_n(A)$ in the descending order from top left. The column vectors of $U$ and $V$ are the left and right eigenvectors of A, as $U=[\bm{u}_{A1},\cdots,\bm{u}_{An}]$ and $V=[\bm{v}_{A1},\cdots,\bm{v}_{An}]$. Note that $\bm{u}_{Ai}$ and $\bm{v}_{Ai}$ are the regular (=right) eigenvectors of $AA^T$ and $A^TA$, respectively. Substituting the maximum eigenvector of $\Phi^T\Phi$, denoted by $\bm{v}(\Phi)$, with $\bm{x}$ in Eq. \eqref{eq:norm-equality}, we obtain
\begin{equation}\label{eq:lb-proof1}
\begin{split}
    \sigma_{\text{max}}(\Phi^TH^TR^{-1}H\Phi)&=\left(\sup_{\|\bm{x}\|\neq\bm{0}}\frac{\|S\Phi\bm{x}\|_2}{\|\bm{x}\|_2}\right)^2\\
    &\geq\|S\Phi\bm{v}(\Phi)\|_2^2=\sigma^2_{\text{max}}(\Phi)\|S\bm{u}(\Phi)\|_2^2=\sigma_{\text{max}}(\Phi\Phi^T)\|S\bm{v}_{\text{CGT}}\|_2^2.
\end{split}
\end{equation}
In the last equality we used $\sigma^2_{\text{max}}(\Phi)=\sigma_{\text{max}}(\Phi\Phi^T)$ and $\bm{u}(\Phi)=\bm{v}(\Phi\Phi^T)\triangleq\bm{v}_{\text{CGT}}$. The alignment $\alpha_i$ of Eq. \eqref{eq:alignment} is the $i$-th element of $V^T_S \bm{v}_{\text{CGT}}$ where $S=U_S D_S V_S^T$. Therefore, the vector inside the squared norm of the last term becomes the linear sum of $\alpha_i\sigma_{S,i}\bm{u}_{S,i}$ as follows:
\begin{equation}\label{eq:lb-proof2}
\begin{split}
\|S\bm{v}_{\text{CGT}}\|_2^2&=\|\alpha_1\sigma_{S,1}\bm{u}_{S,1}+\cdots+\alpha_6\sigma_{S,6}\bm{u}_{S,6}\|_2^2\\
    &= \left(\alpha_1\sigma_{S,1}\bm{u}_{S,1}+\cdots+\alpha_6\sigma_{S,6}\bm{u}_{S,6}\right)^T\left(\alpha_1\sigma_{S,1}\bm{u}_{S,1}+\cdots+\alpha_6\sigma_{S,6}\bm{u}_{S,6}\right)\\
    &=\sum_{i=1}^6 \alpha_i^2\sigma_{S,i}^2=\sum_{i=1}^6 \alpha_i^2\sigma_{i}(H^TR^{-1}H).
\end{split}
\end{equation}
We used the orthogonality of $U$ and $\sigma_{S,i}^2=\sigma_{S^TS,i}=\sigma_{i}(H^TR^{-1}H)$. Combining Eqs. \eqref{eq:lb-proof1} and \eqref{eq:lb-proof2} with $\sigma_5(H^TR^{-1}H)=\sigma_6(H^TR^{-1}H)=0$ completes the proof. 

\section*{Acknowledgments}
The authors gratefully acknowledge support for this research from the Air Force Office of Scientific Research (AFOSR), as part of the Space University Research Initiative (SURI), grant FA9550-22-1-0092 (grant principal investigator: J. Crassidis from University at Buffalo, The State University of New York).

\bibliography{main}

\end{document}